\documentclass[11pt,a4paper]{article}
\usepackage{amssymb,amsmath}
\numberwithin{equation}{section}
\newtheorem{definition}{Definition}[section]
\newtheorem{proposition}[definition]{Proposition}
\newtheorem{lemma}[definition]{Lemma}
\newtheorem{theorem}[definition]{Theorem}
\newtheorem{remark}[definition]{Remark}
\begin{document}
\title{The surface-symmetric Einstein-Vlasov system with cosmological constant}
\author{Sophonie Blaise Tchapnda N. ; Norbert Noutchegueme \\
Department of Mathematics, Faculty of
Science,\\ University of Yaounde I, PO Box 812, Yaounde, Cameroon \\
{e-mail: tchapnda@uycdc.uninet.cm ; nnoutch@uycdc.uninet.cm}}
\date{}
\maketitle
\begin{abstract}
The results on the local existence and continuation criteria
obtained by G. Rein in \cite{rein2} are extended to the case with
a non-zero cosmological constant. It is also shown that for the
spherically symmetric case and a positive cosmological constant
there is a large class of initial data with global existence and
inflationary asymptotics in the future as in the case of plane or
hyperbolic symmetry treated in \cite{tchapnda}. Furthermore we
analyze the behaviour of the energy-momentum tensor at late times.
\end{abstract}
\section{Introduction}

The Einstein-Vlasov system governs the time evolution of a
self-gravitating collisionless gas in the context of general
relativity. In general two classes of initial data are
distinguished. This is the first class : if an isolated body is
studied, the data are called asymptotically flat. Spacetimes that
possess a compact Cauchy hypersurface are called cosmological
spacetimes and data are given on a compact 3-manifold, it is the
second class. In this case the whole universe is modelled and the
"particles" in the kinetic description are galaxies or even
clusters of galaxies. The following is concerned with the second
class.

In \cite{rein2}, G. Rein obtained cosmological solutions of the
Einstein-Vlasov system with surface symmetry written in areal
coordinates. In the present paper we consider the same problem
when a cosmological constant $\Lambda$ is added to the source
terms in the Einstein equations. A motivation for being interested
in this kind of generalization is from the point of view of
astrophysics. In fact present measurements indicate that in our
universe it is the case that $\Lambda > 0$, \cite{straumann}. One
piece of evidence for $\Lambda > 0$ is the data on supernovae of
type Ia, very distant astronomical objects whose distance can be
determined precisely. This shows that the expansion of the
universe is accelerating, which corresponds mathematically to the
fact that the mean curvature tends to a positive constant at late
times. We refer to the original paper \cite{straumann} for the
details of the evidence for $\Lambda > 0$. Another reason for
being interested in the Einstein-Vlasov system with $\Lambda$ is
that having $\Lambda > 0$ makes it easier to obtain statements
about asymptotic behaviour, as shown by the results of
\cite{tchapnda}, the results obtained in the present paper on
spherical symmetry, and those obtained in \cite{lee} for spatially
homogeneous case. Hence from a purely mathematical point of view
it is interesting to study the case $\Lambda > 0$.

The interesting results of \cite{tchapnda} rely essentially on
having suitable theorems on local existence and uniqueness that
are provided in the present paper. Furthermore a new result on
global existence in the spherically symmetric case which uses
these theorems is presented. It reveals a behaviour qualitatively
quite different from that with $\Lambda = 0$ for which it is
proved in \cite{rein2} that no global solution toward the future
could exist.

Now let us formulate the Einstein-Vlasov system with cosmological
constant. All the particles are assumed to have rest mass,
normalized to unity, and to move forward in time so that their
number density $f$ is a non-negative function supported on the
mass shell
\begin{align*}
PM:=
\{g_{\alpha\beta}p^{\alpha}p^{\beta} = -1,
 \ p^0 > 0 \},
\end{align*}
a submanifold of the tangent bundle $TM$ of the space-time
manifold $M$ with metric $g$ of signature $-+++$. We use
coordinates $(t,x^a)$ with zero shift and corresponding canonical
momenta $p^\alpha$ ; Greek indices always run from $0$ to $3$, and
Latin ones from $1$ to $3$. On the mass shell $PM$ the variable
$p^0$ becomes a function of the remaining variables $(t, x^a,
p^b)$ :
\begin{align*}
p^0 = \sqrt{-g^{00}}\sqrt{1+g_{ab}p^{a}p^{b}}.
\end{align*}
The Einstein-Vlasov system now reads
\begin{align*}
\partial_{t}f + \frac{p^{a}}{p^{0}} \partial_{x^{a}}f -
\frac{1}{p^{0}}\Gamma_{\beta\gamma}^{a} p^{\beta} p^{\gamma} \partial_{p^{a}}f
= 0 \\ G_{\alpha\beta} + \Lambda g_{\alpha\beta} = 8 \pi T_{\alpha\beta} \\
T_{\alpha\beta} = - \int_{\mathbb{R}^{3}}f p_{\alpha}
p_{\beta}|g|^{1/2} \frac{dp^{1}dp^{2}dp^{3}}{p_{0}},
\end{align*}
where $p_{\alpha} = g_{\alpha\beta} p^{\beta}$,
$\Gamma_{\beta\gamma}^{\alpha}$ are the Christoffel symbols, $|g|$
denotes the determinant of the metric $g$, $G_{\alpha\beta}$ the
Einstein tensor, $\Lambda$ the cosmological constant, and
$T_{\alpha\beta}$ is the energy-momentum tensor.

Here we adopt the definition of spacetimes with surface symmetry,
i.e., spherical, plane or hyperbolic symmetry given in
\cite{rendall}. We write the system in areal coordinates, i.e.
coordinates are chosen such that $R=t$, where $R$ is the area
radius function on a surface of symmetry. The circumstances under
which coordinates of this type exist are discussed in
\cite{andreasson} for the Einstein-Vlasov system with vanishing
$\Lambda$, and in \cite{tchapnda} for the case with $\Lambda$. In
such coordinates the metric takes the form
\begin{equation} \label{eq:1.1}
  ds^2 = -e^{2\mu(t,r)}dt^2 + e^{2\lambda(t,r)}dr^2 + t^2
  (d\theta^2 + \sin_{k}^{2}\theta d\varphi^{2})
\end{equation}
where
\begin{displaymath}
 \sin_{k}\theta := \left\{ \begin{array}{ll}
\sin\theta & \textrm{if $k=1$}\\
1 & \textrm{if $k=0$}\\
\sinh\theta & \textrm{if $k=-1$}
  \end{array} \right.
\end{displaymath}

Here $t > 0$, the functions $\lambda$ and $\mu$ are periodic in
$r$ with period $1$. It has been shown in \cite{rein1} and
\cite{andreasson} that due to the symmetry $f$ can be written as a
function of
\begin{align*}
t, r, w := e^{\lambda}p^1 \ & \textrm{and} \  F := t^{4}(p^2)^2 +
t^4 \sin_{k}^{2}\theta (p^{3})^{2}, \ \textrm{with} \ r,w \in
\mathbb{R} \ ; \ F \in [0,+\infty[ \ ;
\end{align*}
i.e. $f = f(t, r, w, F)$. In these variables we have $p^0 =
e^{-\mu}\sqrt{1 + w^{2} + F/t^{2}}$. After calculating the Vlasov
equation in these variables, the non-trivial components of the
Einstein tensor, and the energy-momentum tensor and denoting by a
dot or by prime the derivation of the metric components with
respect to $t$ or $r$ respectively, the complete Einstein-Vlasov
system reads as follows :
\begin{equation} \label{eq:1.2}
\partial_{t}f + \frac{e^{\mu-\lambda}w}{\sqrt{1+w^{2}+F/t^{2}}}
\partial_{r}f - (\dot{\lambda}w +
e^{\mu-\lambda}\mu'\sqrt{1+w^{2}+F/t^{2}})\partial_{w}f = 0
\end{equation}
\begin{equation} \label{eq:1.3}
e^{-2\mu} (2t\dot{\lambda}+1)+ k - \Lambda t^{2} = 8 \pi t^{2}\rho
\end{equation}
\begin{equation} \label{eq:1.4}
e^{-2\mu} (2t\dot{\mu}-1)- k + \Lambda t^{2} = 8 \pi t^{2}p
\end{equation}
\begin{equation} \label{eq:1.5}
\mu' = -4 \pi t e^{\lambda+\mu}j
\end{equation}
\begin{equation} \label{eq:1.6}
e^{-2\lambda}\left(\mu'' + \mu'(\mu' - \lambda')\right) -
e^{-2\mu}\left(\ddot{\lambda}+(\dot{\lambda}-
\dot{\mu})(\dot{\lambda}+\frac{1}{t})\right) + \Lambda  = 4 \pi q
\end{equation}
where
\begin{equation} \label{eq:1.7}
\rho(t, r) := \frac{\pi}{t^{2}} \int_{-\infty}^{\infty}
\int_{0}^{\infty} \sqrt{1+w^{2}+F/t^{2}} f(t, r, w, F) dF dw =
e^{-2\mu}T_{00}(t, r)
\end{equation}
\begin{equation} \label{eq:1.8}
p(t, r) := \frac{\pi}{t^{2}} \int_{-\infty}^{\infty}
\int_{0}^{\infty} \frac{w^{2}}{\sqrt{1+w^{2}+F/t^{2}}} f(t, r, w,
F) dF dw = e^{-2\lambda}T_{11}(t, r)
\end{equation}
\begin{equation} \label{eq:1.9}
j(t, r) := \frac{\pi}{t^{2}} \int_{-\infty}^{\infty}
\int_{0}^{\infty} w f(t, r, w, F) dF dw = -e^{\lambda +
\mu}T_{01}(t, r)
\end{equation}
\begin{equation} \label{eq:1.10}
q(t, r) := \frac{\pi}{t^{4}} \int_{-\infty}^{\infty}
\int_{0}^{\infty} \frac{F}{\sqrt{1+w^{2}+F/t^{2}}} f(t, r, w, F)
dF dw = \frac{2}{t^{2}}T_{22}(t, r).
\end{equation}

We study the initial value problem for the Einstein-Vlasov system
(\ref{eq:1.2})-(\ref{eq:1.6}) and prescribe initial data at some
time $t = t_0 > 0$,
\begin{eqnarray*}
f(t_0, r, w, F)= \overset{\circ}{f}(r, w, F), \ \lambda(t_0, r) =
\overset{\circ}{\lambda}(r) , \ \mu(t_0, r) =
\overset{\circ}{\mu}(r).
\end{eqnarray*}

The organization of the paper is as follows. In section $2$ we
give some preliminaries useful to generalize the results obtained
in \cite{rein2} to our situation and we discuss the constraint
equation on the initial data. In section $3$ we adapt to our
situation the results of \cite{rein2} on the local-in-time
existence and uniqueness of solutions in both time directions,
together with a continuation criterion. In section $4$ we prove
that, for $\Lambda < 0$, no solution exists for all $t \geq t_{0}$
by showing that in that case the lapse function $e^{2 \mu}$ blows
up ; next, by using the vacuum solutions we show that in the case
$\Lambda > 0$ and $k=1$ the global solution may exist or not for
all $t \geq t_{0}$, depending on the choice of the initial data.
In section $5$ we prove that, under a suitable condition on the
initial data, the global existence result together with the
asymptotics obtained in \cite{tchapnda} for plane or hyperbolic
symmetry can be generalized to the spherically symmetric case.
Finally section $6$ deals with estimates on the relative sizes of
different components of the energy-momentum tensor which can be
related to physically interesting statements ; our results hold
for the 3 types of symmetry and strengthen some results of
\cite{tchapnda}.

\section{Preliminaries}

Here are the regularity properties which are required for a
solution, the same as in \cite{rein2}.

\begin{definition} \label{def:2.1} Let $I \subset ]0, \infty[$ be
an interval \\
(a) $f \in C^{1}(I \times \mathbb{R}^{2} \times [0, \infty[)$ is
regular, if $f(t, r+1, w, F) = f(t, r, w, F)$ for $(t, r, w, F)
\in I \times \mathbb{R}^{2} \times [0, \infty[$, $f \geq 0$, and
${\rm supp}f(t, r, ., .)$ is compact, uniformly in $r$ and locally
uniformly in $t$.\\
(b) $\rho$ (or $p$, $j$, $q$)$\in C^{1}(I \times \mathbb{R})$ is
regular, if $\rho(t, r+1) = \rho(t, r)$ for $(t, r) \in I \times
\mathbb{R}$ \\
(c) $\lambda \in C^{1}(I \times \mathbb{R})$ is regular, if
$\dot{\lambda} \in C^{1}(I \times \mathbb{R})$ and $\lambda(t,
r+1) = \lambda(t, r)$ for $(t, r) \in I \times
\mathbb{R}$ \\
(d) $\mu\in C^{1}(I \times \mathbb{R})$ is regular, if $\mu' \in
C^{1}(I \times \mathbb{R})$ and $\mu(t, r+1) = \mu(t, r)$ for $(t,
r) \in I \times \mathbb{R}$.
\end{definition}
Such functions are identified with their restrictions to the
interval $[0, 1]$ with respect to $r$.

The following result shows how to obtain $\lambda$ and $\mu$ from
the field equations (\ref{eq:1.2}) and (\ref{eq:1.3}) for given
$\rho$ and $p$. This proposition ends with a statement which
allows us to drop the additional condition
$\overset{\circ}{\mu}(r)<0$ imposed in \cite{rein2} for the case
of hyperbolic symmetry.

\begin{proposition} \label{p:2.2} Let $\rho$ and $p$ :
$I \times \mathbb{R}\longrightarrow
\mathbb{R}$ be regular, $I \subset ]0, \infty[$ an interval with
$t_0 \in I$, $\overset{\circ}{\lambda}, \overset{\circ}{\mu} \in
C^{1}(\mathbb{R})$ with $\overset{\circ}{\lambda}(r) =
\overset{\circ}{\lambda}(r+1)$, $\overset{\circ}{\mu}(r) =
\overset{\circ}{\mu}(r+1)$ for $r \in \mathbb{R}$, and assume that
\begin{equation} \label{eq:2.3}
\frac{t_{0}(e^{-2 \overset{\circ}{\mu}(r)} + k)}{t} - k + \frac{8
\pi}{t}\int_{t}^{t_{0}}s^{2}p(s, r) ds + \frac{\Lambda}{3t}(t^{3}
- t_{0}^{3} ) > 0 , (t,r) \in I \times \mathbb{R}
\end{equation}
Then the equations (\ref{eq:1.2}) and (\ref{eq:1.3}) have unique,
regular solution $(\lambda, \mu)$ on $I \times \mathbb{R}$ with
$\lambda(t_0) = \overset{\circ}{\lambda}$ and $\mu(t_0) =
\overset{\circ}{\mu}$. The solution is given by
\begin{equation} \label{eq:2.4}
e^{-2 \mu(t, r)} = \frac{t_{0}(e^{-2 \overset{\circ}{\mu}(r)} +
k)}{t} - k + \frac{8 \pi}{t}\int_{t}^{t_{0}}s^{2}p(s, r) ds +
\frac{\Lambda}{3t}(t^{3} - t_{0}^{3} )
\end{equation}
\begin{equation} \label{eq:2.5}
\dot{\lambda}(t, r) = 4 \pi t e^{2 \mu(t, r)}\rho(t, r) - \frac{1
+ k e^{2 \mu(t, r)} }{2t} + \frac{\Lambda}{2}t e^{2 \mu(t, r)}
\end{equation}
\begin{equation} \label{eq:2.6}
\lambda(t, r) = \overset{\circ}{\lambda}(r) -
\int_{t}^{t_0}\dot{\lambda}(s, r) ds
\end{equation}
If $I = ]T, t_0]$ (respectively $I = [t_0, T[$) with $T \in [0,
t_0[$ (resp. $T \in ]t_0, \infty]$ ) then there exists some
$T^{\star} \in [T, t_0[$ (resp. $T^{\star} \in ]t_0, T]$) such
that condition (\ref{eq:2.3}) holds on $]T^{\star}, t_0] \times
\mathbb{R}$ (resp. $[t_0, T^{\star}[ \times \mathbb{R}$).
$T^{\star}$ is independent of $p$ for $I = ]T, t_0]$, whereas it
depends on $p$ for $I = [t_0, T[$.
\end{proposition}
{\bf Proof} : The proof for the first part of the present
proposition is the same as that for Proposition 2.4 in
\cite{rein1}. Let us prove the
second part :\\
 If $I = ]T, t_0]$, the function of $t$ and $r$
defined by the right hand side of (\ref{eq:2.4}) is bounded from
below by
\begin{displaymath}
h(t,r) = \frac{t_{0}(e^{-2\overset{\circ}{\mu}} + k)}{t} - k +
\frac{\Lambda}{3t}(t^{3} - t_{0}^{3}).
\end{displaymath}
Since $\overset{\circ}{\mu}$ is continuous and periodic in $r$, it
is bounded, so there exists some $\beta(t_0) > 0$ such that
\begin{displaymath}
h(t_{0},r) = e^{-2\overset{\circ}{\mu}} > \beta(t_0) > 0.
\end{displaymath}
Thus the continuity of $t \mapsto h(t,r)$ at $t = t_0$ implies the
existence of some $T^{\star} \in ]0, t_0[$ such that
\begin{equation} \label{eq:2.7}
h(t,r) > \frac{\beta(t_0)}{2} > 0 \ \ \textrm{for every} \ t \in
]T^{\star}, t_0],
\end{equation}
i.e. (\ref{eq:2.3}) holds for $t \in ]T^{\star}, t_0]$. Now if $I
= [t_0, T[ $, we proceed as above by setting in this case
\begin{displaymath}
h(t,r) = \frac{t_{0}(e^{-2\overset{\circ}{\mu}} + k)}{t} - k -
\frac{8 \pi}{t}\int_{t_0}^{t}s^{2}p(s,r)ds +
\frac{\Lambda}{3t}(t^{3} - t_{0}^{3}) . \ \ \Box
\end{displaymath}
The other preliminary results of \cite{rein2} can be generalized
with minor changes to the case with non-zero $\Lambda$. We have :
\begin{proposition}
1) For $\overset{\circ}{f} \in C^{1}(\mathbb{R}^{2}\times [0,
+\infty[)$ with $\overset{\circ}{f}(r+1,w,F)=
\overset{\circ}{f}(r,w,F)$ and given regular $\lambda$, $\mu$, the
Vlasov equation (\ref{eq:1.2}) has a unique regular solution $f$.
The solution is given by
\begin{equation}\label{eq:2.6}
  f(t,r,w,F) = \overset{\circ}{f}((R,W)(t_0, t, w,F), F)
\end{equation}
where $s \mapsto (R,W)(s)$ is the solution of the characteristic
system associated to (\ref{eq:1.2}) such that
$(R,W)(t,t,r,w,F)=(r,w)$.\\
2) The subsystem (\ref{eq:1.2})-(\ref{eq:1.3})-(\ref{eq:1.4}) is
equivalent to the full system (\ref{eq:1.2})-(\ref{eq:1.6}),
provided that the initial data satisfy (\ref{eq:1.5}) at $t=t_0$.
\end{proposition}

We conclude this section with a remark dealing with the
solvability of the constraint equation (\ref{eq:1.5}) for $t =
t_0$. Note that this result is the same for any value of $\Lambda$
and completes the results of \cite{rein2}.
\begin{remark} \label{r:2.6} The constraint equation $\overset{\circ}{\mu}' =
 -4 \pi t_0 e^{\overset{\circ}{\lambda} +
\overset{\circ}{\mu}}\overset{\circ}{j}$ is solvable.
\end{remark}
{\bf Proof} : Indeed this equation is equivalent to
\begin{eqnarray*}
(e^{-\overset{\circ}{\mu}})' = 4 \pi t_0
e^{\overset{\circ}{\lambda}}\overset{\circ}{j} = \frac{4
\pi^{2}}{t_0}e^{\overset{\circ}{\lambda}} \int_{-\infty}^{\infty}
\int_{0}^{\infty}w \overset{\circ}{f}(r,w,F) dF dw
\end{eqnarray*}
To solve this we need to impose the condition that
\begin{eqnarray*}
I(\overset{\circ}{f}) := \frac{4 \pi^{2}}{t_0}\int_{0}^{1}
\int_{-\infty}^{\infty} \int_{0}^{\infty}
e^{\overset{\circ}{\lambda}}w \overset{\circ}{f}(r,w,F) dF dw dr =
0
\end{eqnarray*}
since $e^{-\overset{\circ}{\mu}(r)}$is periodic in $r$ with period
$1$. Let us choose $\overset{\circ}{\lambda}$ freely, and
$\bar{f}$ a nonnegative function. \\ Firstly if $I(\bar{f}) = 0$
then it suffices to take $\overset{\circ}{f} = \bar{f} $.
\\ Next if $I(\bar{f}) > 0$ then we fix $\Phi \in
C_{c}^{\infty}$, $\Phi \geq 0$ such that $\Phi$ does not vanish
identically and ${\rm supp}\Phi \subset \{ w < 0 \}$, and we set
$\overset{\circ}{f}(r,w,F) = \bar{f}(r,w,F) + a \Phi(r,w,F)$ where
$a$ is a positive constant such that
\begin{align*}
I(\overset{\circ}{f}) & = \frac{4 \pi^{2}}{t_0}\int_{0}^{1}
\int_{-\infty}^{\infty} \int_{0}^{\infty}
e^{\overset{\circ}{\lambda}}w \bar{f}(r,w,F) dF dw dr\\
 & + \frac{4 \pi^{2} a}{t_0}\int_{0}^{1} \int_{-\infty}^{\infty}
\int_{0}^{\infty} e^{\overset{\circ}{\lambda}}w \Phi(r,w,F) dF dw
dr \\
& = 0
\end{align*}
Now if $I(\bar{f}) < 0$ then we proceed in the same way as above
but now with\\ ${\rm supp}\Phi \subset \{ w > 0 \}$.\\ Thus we
determine a candidate for $e^{-\overset{\circ}{\mu}}$ up to an
additive constant, having given $\overset{\circ}{\lambda}$ and
$\bar{f}$ freely. Choosing a suitable constant ensures that this
function is positive and thus of the form
$e^{-\overset{\circ}{\mu}}$. $\Box$

\section{Local existence and continuation of solutions}

This section provides local existence and uniqueness results with
the continuation criteria in both time directions.

\begin{theorem} \label{t:3.1} Let
 $\overset{\circ}{f} \in C^{1}(\mathbb{R}^{2} \times [0, \infty[)$
 with $\overset{\circ}{f}(r+1,w,F) = \overset{\circ}{f}(r,w,F)$
 for $(r,w,F) \in \mathbb{R}^{2} \times [0, \infty[$,
 $\overset{\circ}{f}\geq 0$, and
 \begin{eqnarray*}
 w_{0} := \sup \{ |w| / (r,w,F) \in {\rm supp} \overset{\circ}{f} \} <
 \infty
 \end{eqnarray*}
\begin{eqnarray*}
 F_{0} := \sup \{ F / (r,w,F) \in {\rm supp} \overset{\circ}{f} \} <
 \infty
 \end{eqnarray*}
 Let $\overset{\circ}{\lambda} \in C^{1}(\mathbb{R})$,
 $\overset{\circ}{\mu} \in C^{2}(\mathbb{R})$ with
 $\overset{\circ}{\lambda}(r) = \overset{\circ}{\lambda}(r+1)$,
$\overset{\circ}{\mu}(r) = \overset{\circ}{\mu}(r+1)$ for $r \in
\mathbb{R}$, and
\begin{eqnarray*}
\overset{\circ}{\mu}'(r) =
 -4 \pi t_0 e^{\overset{\circ}{\lambda} +
\overset{\circ}{\mu}}\overset{\circ}{j}(r) = -\frac{4
\pi^{2}}{t_0}e^{\overset{\circ}{\lambda}+ \overset{\circ}{\mu}}
\int_{-\infty}^{\infty} \int_{0}^{\infty}w
\overset{\circ}{f}(r,w,F) dF dw, \ \  r \in \mathbb{R}
\end{eqnarray*}
Then there exists a unique, left maximal, regular solution $(f,
\lambda, \mu)$ of (\ref{eq:1.2})-(\ref{eq:1.6}) with $(f, \lambda,
\mu)(t_0) = (\overset{\circ}{f}, \overset{\circ}{\lambda},
\overset{\circ}{\mu})$ on a time interval $]T, t_0]$ with $T \in
[0, t_0[$.
\end{theorem}
This is the analogue of the first part of theorem 3.1 in
\cite{rein2}

{\bf Proof} : We just indicate the points of the proof which
differ from the proof of theorem 3.1 in \cite{rein2}.\\ The
iterates are defined in the same way except that now
(\ref{eq:2.4}) is used to define $\mu_n$ on the interval $]T_n,
t_0]$, where
\begin{eqnarray*}
 T_{n}  :=  \inf \left\{ t' \in ]T_{n-1}, t_{0}[ /
\frac{t_{0}(e^{-2 \overset{\circ}{\mu}(r) } + k)}{t} - k + \frac{8
\pi}{t}\int_{t}^{t_0}s^{2}p_{n}(s, r) ds \right.{}\\
\left.
{} + \frac{\Lambda}{3t}(t^{3} - t_{0}^{3}) > 0, r \in \mathbb{R}, t
\in [t', t_{0}]\right\},
\end{eqnarray*}
$]T_{n-1}, t_0]$ being the existence interval of the previous
iterates and $T_0 = 0$. By Proposition \ref{p:2.2}, $T_n \leq
T^{\star}$ for all $n$. So all the iterates are well defined and
regular on the fixed time interval $]T^{\star}, t_0]$ (this
interval has the role played by $]0.1]$ in the proof of theorem
3.1 in \cite{rein2}). The other steps of the proof remain the same
although the presence of the cosmological constant $\Lambda$ leads
to some changes. For instance here the constants $c_1$, $c_2$ and
$c_3$ occurring in the estimates are given by :
\begin{eqnarray} \label{eq:3.1}
\begin{cases}
c_1 = c_1(\mu(t_0))= \beta(t_0)/2\\
 c_{2} = c_{2}(\overset{\circ}{f}, F_{0},
\overset{\circ}{\mu}, \Lambda) := \frac{c}{c_1} (1 + 1/c_{1})(1 +
F_{0})^{2}(1 +
\parallel \overset{\circ}{f} \parallel)(\frac{\mid \Lambda
\mid t_{0}^{3}}{2c_{1}} + 1)\\
c_{3} := \parallel e^{-2
\overset{\circ}{\mu}}\overset{\circ}{\mu}' \parallel +
\parallel \overset{\circ}{\lambda}' \parallel + 1 + \mid \Lambda
\mid
\end{cases}
\end{eqnarray}
 The constant $c_1$ is a lower bound of
$e^{-2\mu_{n}(t,r)}$ and is independent of $n$ by Proposition
\ref{p:2.2}. $\Box$

Let us now state the continuation criterion for $t$ decreasing.
\begin{theorem} \label{t:3.2}
Let $(\overset{\circ}{f}, \overset{\circ}{\lambda},
\overset{\circ}{\mu})$ be initial data as in Theorem \ref{t:3.1}.
Assume that $(f, \lambda, \mu)$ is a solution of
(\ref{eq:1.2})-(\ref{eq:1.6}) on a left maximal interval of
existence $]T, t_0]$. If
\begin{eqnarray*}
\sup \{ |w| / (t, r, w, F) \in {\rm supp} f \} < \infty
\end{eqnarray*}
and
\begin{eqnarray*}
\sup \{ e^{2\mu(t, r)} / r \in \mathbb{R}, t\in ]T, t_0] \} <
\infty
\end{eqnarray*}
then $T = 0$.
\end{theorem}
This is the analogue of the second part of theorem 3.1 in
\cite{rein2}.\\ {\bf Proof} : The argument is the same as in {\it
Step} 6 in the proof of theorem 3.1 in \cite{rein2} except that
here, for $t_1 \in ]T,t_0]$, the constants $c_1$, $c_2$, $c_3$ are
defined in the same way as in (\ref{eq:3.1}), considering this
time the data at time $t=t_1$ ; and $c_1(\mu(t_1)) \geq
\beta^{\star}$ with $\frac{1}{\beta^{\star}} = \sup \{ e^{2\mu(t,
r)} / r \in \mathbb{R}, t\in ]T, t_0] \}$. Thus there exists a
constant $c_{2}^{\star}
> 0$ such that
\begin{eqnarray*}
c_{2}(f(t_1), F_0, \mu(t_1), \Lambda)/s^2 \leq c_{2}^{\star}
  \qquad \textrm{for $t_{1} \in ]T, t_0]$ and $s \in [T/2, t_0]$}.
\end{eqnarray*}
The other points remain the same having replaced $t_0$ by $t_1$.
$\Box$

Next we state the analogue of Theorem \ref{t:3.1} and Theorem
\ref{t:3.2} for $t \geq t_0$ which generalizes with minor changes
Theorem 6.1 and 6.2 in \cite{rein2}, to the case with non-zero
$\Lambda$.
\begin{theorem} \label{t:3.3} Let
$(\overset{\circ}{f}, \overset{\circ}{\lambda},
\overset{\circ}{\mu})$ be initial data as in Theorem \ref{t:3.1}.
Then there exists a unique, right maximal, regular solution $(f,
\lambda, \mu)$ of (\ref{eq:1.2})-(\ref{eq:1.6}) with $(f, \lambda,
\mu)(t_0) = (\overset{\circ}{f}, \overset{\circ}{\lambda},
\overset{\circ}{\mu})$ on a time interval $[t_0, T[$ with $T \in
]t_0, \infty]$. If
\begin{eqnarray*}
\sup \{ e^{2\mu(t, r)} / r \in \mathbb{R}, t\in [t_0, T[ \} <
\infty
\end{eqnarray*}
then $T = \infty$.
\end{theorem}

\section{Particular cases for future global existence}

In this section we prove that for $\Lambda < 0$ no solution exists
for all $t \geq t_0$ and for $\Lambda
> 0 $ and $k = 1$, the solution needs not exist for
all $t \geq t_0$.
\begin{proposition} \label{p:4.1} 1) In the case $\Lambda < 0$, no
solution exists for all $t \geq t_0$.\\
2) For $\Lambda > 0$ and $k = 1$, the solution may exist or not
for all $t \geq t_0$, depending on the choice of $t_0$ and of the
initial data.
\end{proposition}
{\bf Proof} 1) If $\Lambda <0$, then for any solution $(f,
\lambda, \mu)$ we have [see (2.2)] :
\begin{equation*}
e^{-2\mu(t, r)}  = \frac{t_{0}(e^{-2 \overset{\circ}{\mu}(r)} + k
)}{t} - k - \frac{8\pi}{t}\int_{t_0}^{t}s^{2}p(s, r) ds +
\frac{\Lambda}{3t}(t^{3} - t_{0}^{3})
\end{equation*}
thus
\begin{equation}\label{eq:4.1}
e^{-2\mu(t, r)} \leq \frac{t_{0}(e^{-2 \overset{\circ}{\mu}(r)} +
k )}{t} - k + \frac{\Lambda}{3t}(t^{3} - t_{0}^{3})
\end{equation}
 has to hold on the interval of existence $[t_0, T[$. Since
the right hand side of this estimate tends to $-\infty$ as $t
\rightarrow \infty$ it follows that $T < \infty$ and \\$\parallel
e^{2\mu(t)}\parallel \longrightarrow \infty$ as $t \rightarrow T$,
by Theorem \ref{t:3.3}.\\
2) For $\Lambda > 0$ and $k = 1$, consider the vacuum case. The
equation (\ref{eq:1.4}) then becomes :
\begin{equation} \label{eq:4.2}
e^{-2\mu}(2t \dot{\mu} - 1) - 1 + \Lambda t^{2} = 0
\end{equation}
(\ref{eq:4.2}) is equivalent to
\begin{equation*}
\partial_{t}(te^{-2\mu}) = \Lambda t^{2} - 1
\end{equation*}
integrating this with respect to $t$ over $[t_0,t]$ yields
\begin{equation} \label{eq:4.3}
e^{-2\mu} = t^{-1}(\frac{\Lambda t^{3}}{3} - t + C), \
\textrm{where} \ C= t_0 e^{-2
\overset{\circ}{\mu}(r)}-\frac{\Lambda t_{0}^{3}}{3}+t_0.
\end{equation}
The solution can be defined only if $\frac{\Lambda t^{3}}{3} - t +
C > 0$. The variations of the function $t \mapsto \frac{\Lambda
t^{3}}{3} - t + C$ allow us to conclude that :\\ a) for $t_0 \leq
1/\sqrt{\Lambda}$, the solution exists on the whole interval
$[t_0, +\infty[$ if\\ $C-\frac{2}{3\sqrt{\Lambda}}>0$ whereas if
$C-\frac{2}{3\sqrt{\Lambda}}<0$, the solution exists on some
interval $[t_0, t_1]$ and on some interval $[t_2, +\infty[$,
$0<t_1<t_2$ ;\\ b)  for $t_0>1/\sqrt{\Lambda}$, the solution
exists on $[t_0, +\infty[$ regardless of the sign of $C -
\frac{2}{3\sqrt{\Lambda}}$. $\Box$

Hence, depending on the choice of $t_0$, (\ref{eq:4.3}) shows that
in the case $\Lambda >0$ and $k=1$ there exists a class of initial
data for which global existence fails and there is also a class of
data with global existence in the future. In the next section we
identify a suitable condition on the initial data useful to prove
the latter statement. Note that the result stated for $\Lambda<0$
was obtained in \cite{rein2} in the case $\Lambda=0$, $k=1$, by
using (\ref{eq:4.1}) without the term in $\Lambda$.

\section{The spherically symmetric case}
Let us prove that choosing the initial time $t_0$ to satisfy the
condition $t_0^2 > 1/\Lambda$ is enough to get global existence in
the future and the asymptotic behaviour as in the case $k \le 0$
obtained in \cite{tchapnda}.
\subsection{Global existence in the future}

We write the proof of the analogue of Theorem 2.2 in
\cite{tchapnda}.

We establish a series of estimates which will result in an upper
bound on $\mu$ and will therefore prove that $T = \infty$, using
theorem 3.3. Similar computations were used in \cite{tchapnda}.
Unless otherwise specified in what follows constants denoted by
$C$ will be positive, may depend on the initial data and on
$\Lambda$ and may change their value from line to line.

Let us suppose that $t_0^2 > 1/\Lambda$. Firstly, we have, using
(\ref{eq:2.4}) for $k=1$ :
\begin{align} \label{eq:5.1}
e^{2 \mu(t, r)} & = \left[\frac{t_{0}(e^{-2
\overset{\circ}{\mu}(r)} + 1)}{t} - 1 - \frac{8
\pi}{t}\int_{t_0}^{t}s^{2}p(s, r) ds + \frac{\Lambda}{3t}(t^{3} -
t_{0}^{3} )\right]^{-1} \nonumber\\
& \geq  \frac{t}{C+\frac{\Lambda}{3}t^{3}}, \ t \in [t_0, T[.
\end{align}
In this inequality, $C$ does not depend on $\Lambda$. Let us show
that
\begin{equation}\label{eq:5.2}
\int_{0}^{1} e^{\mu+\lambda}\rho(t,r) dr \leq C t^{-1}, \ t \in
[t_0, T[.
\end{equation}
Using (\ref{eq:1.7}), (\ref{eq:1.8}), (\ref{eq:1.10}), a direct
computation shows that
\begin{equation}\label{eq:5.3}
\frac{d}{dt}\int_{0}^{1} e^{\mu+\lambda}\rho(t,r) dr =
-\frac{1}{t}\int_{0}^{1} e^{\mu+\lambda}\left[2\rho + q
-\frac{\rho+p}{2}(1+e^{2\mu}-\Lambda t^{2} e^{2\mu})\right] dr.
\end{equation}
 (\ref{eq:5.3}) implies the
following, since $q \geq 0$ :
\begin{align*}
\frac{d}{dt}\int_{0}^{1} e^{\mu+\lambda}\rho(t,r) dr & \leq
-\frac{2}{t}\int_{0}^{1} e^{\mu+\lambda}\rho dr +
\frac{1}{t}\int_{0}^{1}e^{\mu+\lambda}\frac{\rho+p}{2}dr
\nonumber\\
& {} + \frac{1}{t}\int_{0}^{1}\frac{(1-\Lambda
t^{2})e^{2\mu}e^{\mu+\lambda}}{2}(\rho +p)dr \nonumber\\
& \leq -\frac{1}{t}\int_{0}^{1} e^{\mu+\lambda}\rho dr,
\end{align*}
we have used the fact that $\rho \geq p$ and $1-\Lambda t^{2} \leq
0$. Integrating over $[t_0,t]$ we obtain by Gronwall's inequality
\begin{align*}
\int_{0}^{1} e^{\mu+\lambda}\rho(t,r) dr & \leq Ct^{-1}
\end{align*}
that is (\ref{eq:5.2}) holds. Using (\ref{eq:5.2}) and the
equation $\mu' = -4\pi te^{\mu+\lambda}j$ we find
\begin{align}\label{eq:5.5}
\mid \mu(t,r)- \int_{0}^{1}\mu(t,\sigma)d\sigma \mid & =
 \mid \int_{0}^{1}\int_{\sigma}^{r}\mu'(t,\tau)d\tau d\sigma \mid
 \leq \int_{0}^{1}\int_{0}^{1}|\mu'(t,\tau)|d\tau d\sigma
 \nonumber\\
 & \leq 4\pi t\int_{0}^{1}e^{\mu+\lambda}|j(t,\tau)|d\tau
 \leq 4\pi t\int_{0}^{1}e^{\mu+\lambda}\rho(t,\tau)d\tau
 \nonumber\\
 & \leq C, \ t \in [t_0, T[, \ r \in [0,1].
\end{align}
Next we show that
\begin{equation}\label{eq:5.6}
  e^{\mu(t,r)-\lambda(t,r)} \leq Ct^{-2}, \ t \in [t_0, T[, \ r
  \in [0,1]
\end{equation}
To see this, observe that by equations (1.3), (1.4) with $k=1$ and
(\ref{eq:5.1})
\begin{align*}
 \frac{ \partial}{\partial t}e^{\mu-\lambda} & =
 e^{\mu-\lambda}\left[4\pi
te^{2\mu}(p-\rho)+\frac{1+e^{2\mu}}{t}-\Lambda t
 e^{2\mu}\right]\leq e^{\mu-\lambda}\left[\frac{1}{t}+\frac{(1-\Lambda t^{2})}{t}
 e^{2\mu}\right]\\
 & \leq \left[ \frac{1}{t} + \frac{1-\Lambda
 t^{2}}{C+\frac{\Lambda}{3}t^{3}}\right]e^{\mu-\lambda}\\
 & \le \left[ \frac{1}{t} - \frac{\Lambda
 t^{2}}{C+\frac{\Lambda}{3}t^{3}}+
 \frac{3}{\Lambda}t^{-3}\right]e^{\mu-\lambda}.
\end{align*}
Integrating this inequality with respect to $t$ over $[t_0,t]$
yields, by Gronwall's lemma :
\begin{equation*}
e^{\mu-\lambda} \leq C \frac{t}{C+\frac{\Lambda}{3}t^{3}} \leq
Ct^{-2},
\end{equation*}
i.e. (\ref{eq:5.6}).

We now estimate the average of $\mu$ over the interval $[0,1]$
which in combination with (\ref{eq:5.5}) will yield the desired
upper bound on $\mu$. We have, using (\ref{eq:1.4}),
(\ref{eq:5.1}), (\ref{eq:5.2}), (\ref{eq:5.6}) and the fact that
$p \leq \rho$ :
\begin{align*}
\int_{0}^{1}\mu(t,r) dr & = \int_{0}^{1}\overset{\circ}{\mu}(r) dr
+ \int_{t_0}^{t}\int_{0}^{1}\dot{\mu}(s,r) dr ds\\
& \leq C+\int_{t_0}^{t}\frac{1}{2s}\int_{0}^{1}[e^{2\mu}(8\pi
s^{2}p + 1-\Lambda s^{2})+1]dr ds\\
& \le C+
\frac{1}{2}\ln(t/t_0)+\frac{1}{2}\int_{t_0}^{t}\frac{1-\Lambda
s^{2} }{C+\frac{\Lambda}{3}s^{3}}ds+4 \pi \int_{t_0}^{t}s
\int_{0}^{1}e^{\mu-\lambda}e^{\mu+\lambda}\rho dr
ds\\
& \leq
C+\frac{1}{2}\ln(t/t_{0})+C\int_{t_0}^{t}s^{-2}ds+C\int_{t_0}^{t}s^{-3}ds
-\frac{1}{2}\int_{t_0}^{t}\frac{\Lambda
s^{2} }{C+\frac{\Lambda}{3}s^{3}}ds\\
& \leq C + \frac{1}{2}\left[\ln
\frac{s}{C+\frac{\Lambda}{3}s^{3}}\right]_{s=t_0}^{s=t}
\end{align*}
 With (\ref{eq:5.5}) this implies
\begin{equation}\label{eq:5.7}
\mu(t,r) \leq C(1+\ln t^{-2}) \leq C, \ t \in [t_0, T[, \ r \in
[0,1]
\end{equation}
which by Theorem 3.3 implies $T = \infty$. Thus we have proven :
\begin{theorem} \label{t:5.1} For initial data as in Theorem
3.1 with $t_0^2 > 1/\Lambda$, the solution of the Einstein-Vlasov
system with positive cosmological constant and spherical symmetry,
written in areal coordinates, exists for all $t \in [t_0, \infty[$
where $t$ denotes the area radius of the surfaces of symmetry of
the induced spacetime. The solution satisfies the estimates
(\ref{eq:5.2}), (\ref{eq:5.6}) and (\ref{eq:5.7}).
\end{theorem}

\subsection{On future asymptotic behaviour}

Under the hypothesis $t_{0}^{2}>1/\Lambda$, we obtain for the
spherically symmetric case analogous results about the asymptotic
behaviour at late times, obtained in \cite{tchapnda} for the case
of plane or hyperbolic symmetry. We obtain with minor changes in
the proofs the analogue of Theorems 3.2 and 3.3 in
\cite{tchapnda}.
\begin{theorem}\label{t:5.2} Consider initial data with spherical
 symmetry for the Einstein-Vlasov system with positive
cosmological constant satisfying the regularity properties
required in Theorem 3.1 with $t_0^2>1/\Lambda$.\\1) If the
gradient of $R$ is initially past-pointing then there is a
corresponding Cauchy development which is future geodesically
complete.\\ 2) Consider a solution $(f,\lambda, \mu)$ of the
Einstein-Vlasov system with spherical symmetry and $\Lambda
>0$ given in the expanding direction. Then the following
properties hold at late times :
\begin{equation}\label{eq:5.8}
  \dot{\lambda} = t^{-1}\left(1+O(t^{-2})\right),
\end{equation}
\begin{equation}\label{eq:5.9}
 \lambda = \ln t \left[1+O\left((\ln t)^{-1}\right)\right],
\end{equation}
\begin{equation}\label{eq:5.10}
  \dot{\mu} = -t^{-1}\left(1+O(t^{-2})\right),
\end{equation}
\begin{equation}\label{eq:5.11}
\mu = -\ln t\left[1+O\left((\ln t)^{-1}\right)\right],
\end{equation}
\begin{equation}\label{eq:5.12}
\mu' = O(t^{-3+2\varepsilon}),
\end{equation}
 with
$\varepsilon \in ]0,2/3[$ ; and
\begin{displaymath}
\lim_{t\rightarrow + \infty} \frac{K_{1}^{1}(t,r)}{K(t,r)} =
\lim_{t\rightarrow + \infty} \frac{K_{2}^{2}(t,r)}{K(t,r)} =
\lim_{t\rightarrow + \infty} \frac{K_{3}^{3}(t,r)}{K(t,r)} =
\frac{1}{3},
\end{displaymath}
where $K(t,r) = K_{i}^{i}(t,r)$ is the trace of the second
fundamental form $K_{ij}$ of the metric. Here $O$ stands for
Landau's notation.
\end{theorem}
\section{Asymptotics of matter terms}
In this section we determine the explicit leading behaviour of the
components $\rho$, $p$, $j$ and $q$ of the energy-momentum tensor
and later on we compare $\rho$ to other matter terms. Note that
these results hold for the spherical, plane or hyperbolic
symmetry. The hypotheses on the data are those required in theorem
3.1.

We first establish the following useful result :
\begin{lemma} \label{l:6.1} For any characteristic $(r,w,F)$, for any
solution of Einstein-Vlasov system with positive cosmological
constant and spherical, plane or hyperbolic symmetry written in
areal coordinates, with initial data as in theorem 3.1 and with
$t_0^2 > 1/\Lambda$ in the case of spherical symmetry, consider
the quantity $u=tw$. Then $u$ converges to a constant along the
characteristics, as $t$ tends to infinity.
\end{lemma}
{\bf Proof} \  For a characteristic $s\mapsto(r,w,F)(s)$ of the
Vlasov equation (\ref{eq:1.2}), we have, using the field equations
(\ref{eq:1.3})-(\ref{eq:1.5}) and the expressions (\ref{eq:1.7}),
(\ref{eq:1.9}) of $\rho$ and $j$ :
\begin{align}\label{eq:6.1}
\dot{w} & = \frac{4
\pi^{2}}{t}e^{2\mu}\int_{-\infty}^{\infty}\int_{0}^{\infty}
\left(\tilde{w}\sqrt{1+w^{2}+F/t^{2}}-w\sqrt{1+\tilde{w}^{2}+\tilde{F}/t^{2}}\right)f
d\tilde{F}d\tilde{w} \nonumber\\
& +\frac{1+ke^{2\mu}}{2t}w -\frac{\Lambda}{2}twe^{2\mu}.
\end{align}
Then $u$ satisfies an equation of the form $\dot u=au+b$ with
\begin{equation*}
a(t)=\frac{3+(k- \Lambda t^2)e^{2\mu}}{2t}
\end{equation*}
and
\begin{equation}
b(t) = 4 \pi^{2}e^{2\mu}\int_{-\infty}^{\infty}\int_{0}^{\infty}
\left(\tilde{w}\sqrt{1+w^{2}+F/t^{2}}-w\sqrt{1+\tilde{w}^{2}+\tilde{F}/t^{2}}\right)f
d\tilde{F}d\tilde{w}
\end{equation}
It suffices to prove that the functions $a$ and $b$ are integrable
up to $t=\infty$ in order to conclude that $u$ does converge to a
limit for large $t$.

Now (2.2) implies
\begin{align*}
e^{2 \mu(t, r)}  \geq  \frac{t}{C-kt+\frac{\Lambda}{3}t^{3}}, \ k
\leq 0
\end{align*}
Using this inequality for $k \leq 0$, inequality (5.1) for $k =1$
and the fact that $k -\Lambda
t^2 \leq 0$ for large $t$ we obtain :\\
$a(t) \leq (\frac{9C}{2\Lambda}t^{-1}-\frac{3k}{2\Lambda})t^{-3}$
for $k \leq 0$ and $a(t) \leq
(\frac{9C}{2\Lambda}t^{-1}+\frac{3}{2\Lambda})t^{-3}$ for $k=1$.\\
Either way, since $t^{-1} \leq t_{0}^{-1}$, $a(t)$ is bounded by a
constant $C$ times by $t^{-3}$ and hence is integrable up to
$t=\infty$.

Now about $b(t)$, by proposition 3.1 of \cite{tchapnda}, the
factor of $4 \pi^{2}e^{2\mu}$ in (6.2) is bounded. So $b(t)$ will
be integrable if $e^{2\mu}$ is integrable. But by (5.10)
$e^{2\mu}$ falls off faster than $t^{-2}$ at late times. Thus
$e^{2\mu}$ is integrable up to $t=+\infty$ and so is $b(t)$. This
completes the proof of Lemma 6.1. $\Box$

This result allows us to obtain estimates stronger than those in
\cite{tchapnda} for the components of the energy-momentum tensor
using the same procedure as in proposition 6 of \cite{lee} :
\begin{proposition} \label{p:6.2}
Under the same hypotheses as in lemma 6.1, the following
properties hold at late times :
\begin{equation} \label{eq:6.2}
  \rho=O(t^{-3}) \ ; \ p=O(t^{-5}) \ ; \ j=O(t^{-4}) \ ; \ q=O(t^{-5})
\end{equation}
\begin{equation} \label{eq:6.3}
  \frac{p}{\rho}=O(t^{-2}) \ ; \ \frac{j}{\rho}=O(t^{-1}) \ ; \ \frac{q}{\rho}=O(t^{-2})
\end{equation}
\end{proposition}
{\bf Proof} \ Lemma 6.1 implies that $u(t)$ is uniformly bounded
in large time $t$. So, using (2.6) and the fact that
$f(t_0,r,w,F)$ has compact support on $w$, there is a constant $C$
such that
\begin{equation}\label{eq:6.4}
  |w| \leq Ct^{-1} \ \textrm{and} \ f(t,r,w,F) =0, \ \textrm{if} \ |w| \geq
  Ct^{-1}.
\end{equation}
Now by (1.7) we have
\begin{equation*}
\rho(t, r) := \frac{\pi}{t^{2}} \int_{-\infty}^{\infty}
\int_{0}^{\infty} \sqrt{1+w^{2}+F/t^{2}} f(t, r, w, F) dF dw ;
\end{equation*}
using (\ref{eq:6.4}) this implies that
\begin{equation*}
\rho(t, r) = \frac{\pi}{t^{2}} \int_{|w| \leq Ct^{-1}}
\int_{0}^{F_0} \sqrt{1+w^{2}+F/t^{2}} f(t, r, w, F) dF dw ;
\end{equation*}
and since $f(t,r,w,F)$ is constant along the characteristics we
deduce from the latter equation that
\begin{equation}\label{eq:6.5}
\rho \leq C t^{-3}.
\end{equation}
Next by (1.8) we have
\begin{equation*}
p(t, r) := \frac{\pi}{t^{2}} \int_{-\infty}^{\infty}
\int_{0}^{\infty} \frac{w^{2}}{\sqrt{1+w^{2}+F/t^{2}}} f(t, r, w,
F) dF dw
\end{equation*}
This implies that
\begin{equation*}
p(t, r) \leq \frac{\pi}{t^{2}} \int_{|w| \leq Ct^{-1}}
\int_{0}^{F_0} w^{2} f(t, r, w, F) dF dw ;
\end{equation*}
thus
\begin{equation}\label{eq:6.6}
p \leq C t^{-5}.
\end{equation}
By (1.9) we have
\begin{equation*}
j(t, r) := \frac{\pi}{t^{2}} \int_{-\infty}^{\infty}
\int_{0}^{\infty} w f(t, r, w, F) dF dw ;
\end{equation*}
thus
\begin{equation}\label{eq:6.7}
j \leq C t^{-4}.
\end{equation}
By (1.10),
\begin{equation*}
q(t, r) := \frac{\pi}{t^{4}} \int_{-\infty}^{\infty}
\int_{0}^{\infty} \frac{F}{\sqrt{1+w^{2}+F/t^{2}}} f(t, r, w, F)
dF dw ;
\end{equation*}
this implies that
\begin{equation}\label{eq:6.8}
q \leq C t^{-5},
\end{equation}
and (6.3) is proved.

 Now let us prove (6.4). We have, using $w^{2}\leq Ct^{-2}$ and $1\leq
 \sqrt{1+w^{2}+F/t^{2}}$:
\begin{eqnarray*}
\frac{p}{\rho} \leq \frac{\int_{|w| \leq Ct^{-1}} \int_{0}^{F_0}
w^{2} f(t, r, w, F) dF dw}{\int_{|w| \leq Ct^{-1}} \int_{0}^{F_0}
\sqrt{1+w^{2}+F/t^{2}} f(t, r, w, F) dF dw} \leq C t^{-2}.
\end{eqnarray*}
Similarly we get
\begin{eqnarray*}
\frac{j}{\rho} \leq C t^{-1} \ ; \ \frac{q}{\rho} \leq C t^{-2}. \
\Box
\end{eqnarray*}
This proposition shows that all other components of the
energy-momentum tensor become negligible with respect to $\rho$.
This has an interpretation that the asymptotics is "dust-like"
(pressure negligible with respect to density) and that "tilt is
asymptotically negligible".

\vskip 10pt\noindent \textbf{Acknowledgements} : The authors thank
A.D. Rendall for helpful comments and suggestions. They
acknowledge support by a research grant from the
VolkswagenStiftung, Federal Republic of Germany.


\begin{thebibliography}{9}
\bibitem{andreasson}
H. Andr\'easson, G. Rein, A.D. Rendall, {\it On the
Einstein-Vlasov system with hyperbolic symmetry}, Math. Proc.
Camb. Phil. Soc. {\bf 134}, (2003), 529-549.
\bibitem{lee}
H. Lee, {\it Asymptotic behaviour of the Einstein-Vlasov system
with a positive cosmological constant}, preprint 2003,
gr-qc/0308035, to appear in Math. Proc. Camb. Phil. Soc.
\bibitem{rein1}
G. Rein, {\it The Vlasov-Einstein system with surface symmetry},
Habilitationsschrift zur Erlangung der venia legendi f\"ur das
Fach Mathematik am Fachbereich Mathematik der
Ludwig-Maximilians-Universit\"at, (Im M\"arz 1995).
\bibitem{rein2}
G. Rein, {\it Cosmological solutions of the Vlasov-Einstein system
with spherical, plane, and hyperbolic symmetry}, Math. Proc. Camb.
Phil. Soc. {\bf 119}, (1996), 739-762.
\bibitem{rendall}
A.D. Rendall, {\it Crushing singularities in spacetimes with
spherical, plane and hyperbolic symmetry}, Class. Quantum Grav.
{\bf 12}, (1995), 1517-1533.
\bibitem{straumann}
N. Straumann, {\it On the cosmological constant problems and the
astronomical evidence for a homogeneous energy density with
negative pressure}, preprint 2002, astro-ph/0203330.
\bibitem{tchapnda}
S.B. Tchapnda and A.D. Rendall, {\it Global existence and
asymptotic behaviour in the future for the Einstein-Vlasov system
with positive cosmological constant}, Class. Quantum Grav. {\bf
20}, (2003), 3037-3049.
\end{thebibliography}
\end{document}